# Blind source separation of baseband RF communication signals using mixed-signal matrix multiplication circuit

B. Madhavan, E. Lee, J. Zusman and A. F. J. Levi

An 8 × 8 mixed-signal matrix multiplier architecture based on 64 hybrid capacitor-resistor multiplying digital to analogue converters implemented in a 65 nm CMOS technology was developed for the application of blind source separation of baseband RF signals. The integrated circuit has 13-bit resolution for each matrix weight and achieves a measured dynamic range of > 62 dB with a bandwidth of > 15 MHz and typical power dissipation of < 30 mW per matrix row. Separation of single-tone signal is measured to be better than 57 dBc.

*Introduction:* Matrix multiplication implemented in mixed-signal circuits can provide a resource to reduce conventional analogue-to-digital conversion (ADC) requirements and offload subsequent digital signal processing (DSP). Applications include signal processing [1], machine learning [2-3], and optimization of real-time signals. For example, linear algebra operations that unmix signals of interest *before* digitization enables use of lower linearity ADCs and reduces DSP. The key mixed-signal circuit block is a programmable matrix multiplier, often realized in a switched-capacitor architecture. However, previously such architectures [1-3] have been implemented for relatively low-resolution and low-speed applications. We show that relatively high resolution, linearity, speed, and low power may be achieved simultaneously in a mixed-signal matrix multiplier designed for blind source separation of four unknown complex communication signals at a 4-antenna RF receiver. We report on a high resolution (13-bit matrix weights) and high dynamic-range (> 62 dB) 8 × 8 mixed-signal matrix-multiplier that has been designed, fabricated, and measured.

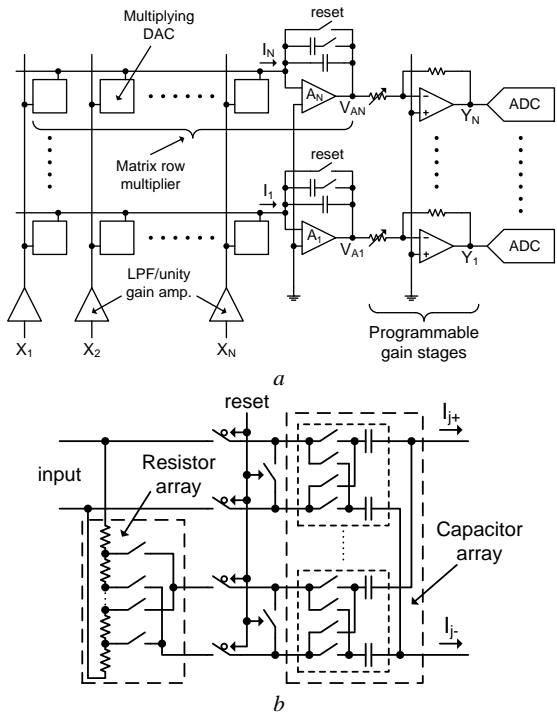

**Fig. 1** *Block diagram of the chip. X is four I and four Q differential analogue inputs. Y is the output.*
*a* Architecture of the mixed-signal matrix multiplier
*b* Detail of resistive and capacitive arrays

*Architecture:* Fig. 1a shows the architecture of the mixed-signal matrix multiplier, which consists of 64 multiplying digital to analogue converters (DACs). Hybrid capacitor-resistor multiplying DACs (Fig. 1b) instead of binary-weighted switched-capacitor multiplying DACs [4-5] are used to achieve high resolution (13-bit) with guaranteed monotonicity and reasonable die area. The thermometer-coded capacitor array consists of 64 pairs of unit capacitors and is determined by the 6 most significant bits (MSBs) plus 1 sign bit. One pair of the unit capacitors obtains an attenuated input signal from the outputs of the 6 least significant bits (LSBs) controlled thermometer-coded resistor array.

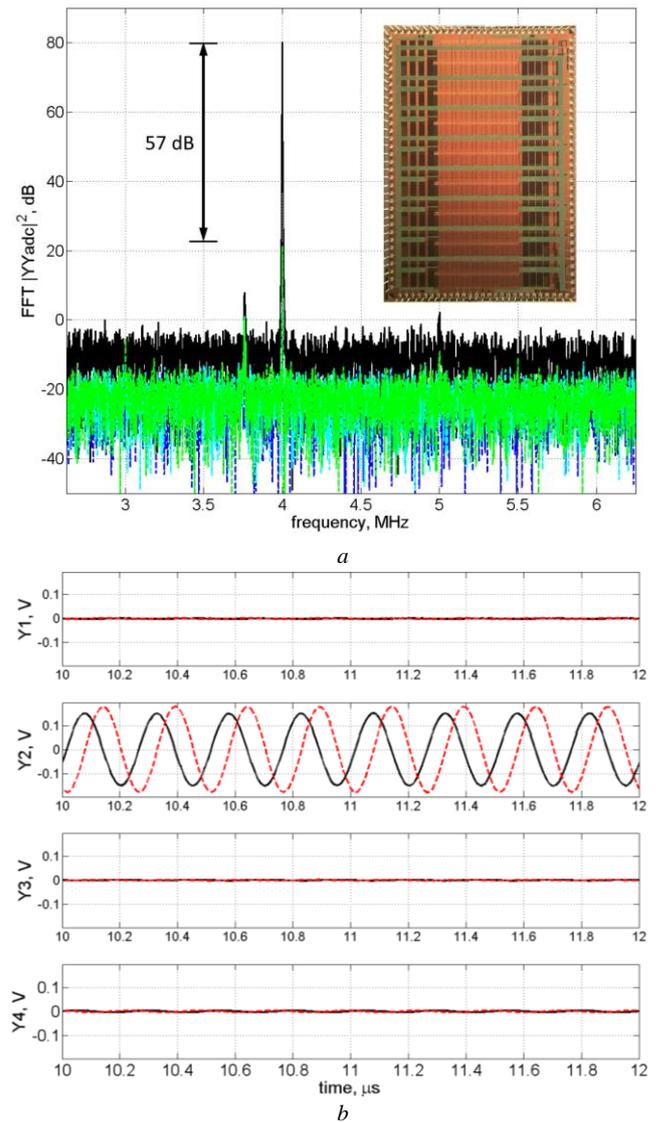

**Fig. 2** *Unmixed measurement of single tone at 4.0 MHz with -10 dBm (0.2 $V_{pp}$) of input power, complex output channel Y2 signal is 72 dBc above frequency spurs. The noise floor of the extracted signal is higher than the other channels due to 18 dB of output gain. Inset is photograph of chip implemented in GF65LPe with area 3.8 mm × 5.5 mm.*
*a* Measured frequency domain power spectra of each complex output channel, $Y_1$, $Y_2$, $Y_3$, and $Y_4$.
*b* Time domain plot of each output channel signal

The multiplying DAC is guaranteed to have 13-bit monotonicity when the 6-bit MSB capacitor array has better than 7-bit accuracy. The minimum time to update the matrix weight and control register is 3.564 μs. The multiplying DACs are reset for 256 ns during loading of matrix weights. After reset, the matrix row multiplier is operated as a continuous-time circuit. This eliminates the need for a high frequency clock signal required in sampled-data architectures before digitization. The feedback capacitor in the matrix row multiplier (Fig. 1*a*) can be switched to a lower value and provide a gain of 6 dB. Each matrix row multiplier output is further amplified by a subsequent 0 – 24 dB (6 dB step) programmable gain stage. Eight six-pole continuous-time filters with cut-off frequency at 6.5 MHz are included at the inputs of the matrix multiplier to suppress unwanted high-frequency signals. The filter is realized by a cascade of three biquad filters. A unity-gain amplifier can be switched in to replace the filter for higher input frequencies in a pass-through mode. The matrix multiplier was implemented in Global Foundries GF65LPe 65 nm CMOS process with a die area of 3.8 mm × 5.5 mm. The chip is designed (and confirmed by measurement) to be calibration free, linear to better than 10-bits, matched across channels better than 1%, handle input signals of up to 1 $V_{pp}$, and have worst-case noise floor less than 0.3 $mV_{RMS}$ (0.15 $mV_{RMS}$ in nominal corner) across process variation and temperature (PVT).

The matrix multiplier may be used to separate linear mixtures of unknown signals, which can be represented in matrix form as **X = AS**, where **S** is the vector for the independent source signals, **A** is the mixing matrix, and **X** is the measured input signal vector after the source signals have been mixed by **A**. Since only **X** is known, the objective is to find both **A** and **S** using minimal information. In a 4-antenna system, the baseband signals are converted into four in-phase (I) and four quadrature (Q) signals, which form the input signal vector **X**. For sub-Gaussian communication waveform statistics, infomax gradient-descent methods [6] can be used to estimate an unmixing matrix **W**, which approximates the exact unmixing matrix $A^{-1}$. Hence, the vector **Y = WX** approximates the separated independent source signal vector **S**.

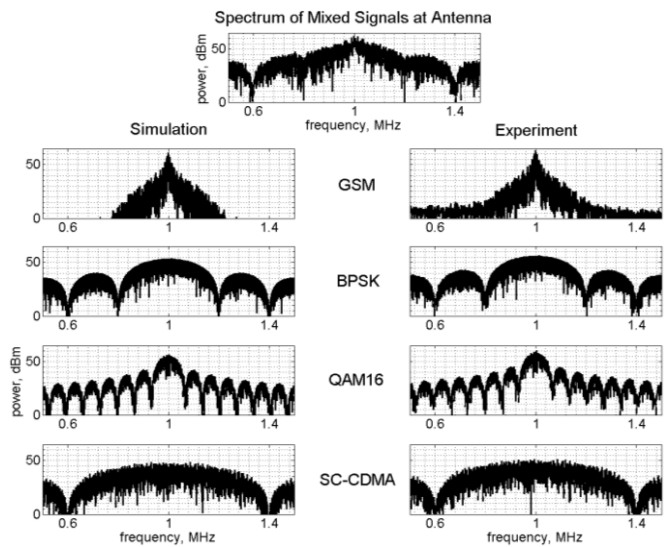

**Fig. 3** *Unmixed four complex communication waveforms (GSM, BPSK, QAM16, and SC-CDMA) each with centre frequency 1 MHz, -24 dBm (40 $mV_{pp}$) input power, and 24 dB of output gain.*
Frequency domain plot of ideal (left) and measured (right) results

*Measurement results:* Laboratory measurements of the chip were performed with arbitrary waveform generators (AWGs) to provide input signals. Fig. 2 shows the experimental results of using the chip to unmix a single I and Q complex tone of frequency 4 MHz from four complex (four I and four Q) mixed inputs with -10 dBm of power on each of the eight (4 × I + 4 × Q = 8) inputs. Stochastic gradient-descent is used to find the weights of matrix **W** that separates the multiple input signals **X** into a single, independent, complex output-signal. Fig. 2*a* is the measured output power-spectra of each of the four complex (I and Q) channels and Fig. 2*b* is the time-domain output signals **Y**, showing separation of the single tone into channel $Y_2$. The complex output signal in channel $Y_2$ is more than 57 dBc above any other signal in complex channels $Y_1$, $Y_3$, or $Y_4$.

The measured pass-through (no input filter) -3 dB bandwidth is 15.4 MHz per channel (30.8 MHz per complex signal). There is less than 0.1 dB compression at 1 MHz with 4 dBm (1 $V_{pp}$) input power and more than 63 dBc IM3 with tones at 1 MHz and 1.1 MHz and 4 dBm input power. Extensive measurements confirm that linearity design goals were realized.

To demonstrate that the chip may be used to blind source separate complex communication signals, the AWG is programmed to generate the corresponding mixed input waveforms. The top of Fig. 3 shows the ideal mixed input power spectrum at **X** consisting of four communication signals with the *same* baseband centre frequency of 1 MHz but with different angle of arrival. The communication signals are GSM, BPSK, QAM16, and SC-CDMA. The left-hand-side of Fig. 3 shows the ideal unmixed spectrum and the right-hand-side shows the experimentally measured spectrum after unmixing by the chip. As may be seen, excellent separation of signals is achieved.

*Conclusion*: A programmable analogue matrix-multiplier designed in 65 nm CMOS technology has been successfully implemented and used to demonstrate blind source separation of baseband RF communication signals. The mixed-signal chip opens a system trade-space where linear algebra operations may be performed prior to conventional receiver ADC and DSP. Beyond RF receiver applications, the programmability, low noise, and high linearity of the chip suggests a versatile generic circuit building block for a diverse range of application domains involving manipulation of real-time data using matrix operations.

*Acknowledgments:* This work was supported by DARPA and SAIC/Leidos.

B. Madhavan, E. Lee, J. Zusman, and A. F. J. Levi
University of Southern California, 3620 South Vermont Avenue, KAP 132, Los Angeles, CA, USA, 90089
Email: madhavan@usc.edu

September 27, 2016.